\documentclass[a4paper, 10pt]{article}

\usepackage{amsmath,amssymb}
\numberwithin{equation}{section}

\usepackage{cite}

\newcommand{\p}{\partial}
\newcommand{\e}{\varepsilon}

\newcommand{\NN}{\mathbb{N}}
\newcommand{\ZZ}{\mathbb{Z}}

\newcommand{\RR}{\mathbb{R}}
\newcommand{\CC}{\mathbb{C}}

\newcommand{\uu}{\mathrm{u}}
\newcommand{\dd}{\mathrm{d}}
\newcommand{\LL}{\Lambda}

\newcommand{\A}{\mathcal{A}}
\newcommand{\E}{\mathcal{E}}

\newtheorem{dfn}{Definition}[section]
\newtheorem{lem}[dfn]{Lemma}

\newtheorem{thm}[dfn]{Theorem}

\newtheorem{emp}[dfn]{Example}
\newtheorem{cnj}[dfn]{Conjecture}

\newenvironment{prfn}[1]{\noindent {\it Proof of #1} \ }{\hfill $\Box$}

\DeclareMathOperator*{\res}{\mathrm{res}}

\begin{document}

\title{On Tautological Flows of \\ Partial Difference Equations}
\author{Zhonglun Cao \and Si-Qi Liu \and Youjin Zhang} 
\date{}
\maketitle

\begin{abstract}
    We propose a new analyzing method, which is called the tautological flow method, to analyze the integrability of
    partial difference equations (P$\Delta$Es) based on that of partial differential equations (PDEs).
    By using this method, we prove that the discrete $q$-KdV equation is a discrete symmetry of the $q$-deformed KdV
    hierarchy and its bihamiltonian structure, and we also demonstrate how to directly search for continuous symmetries
    and bihamiltonian structures of P$\Delta$Es by using the approximated tautological flows and their quasi-triviality
    transformation.
\end{abstract}

\tableofcontents

\section{Introduction}

A partial difference equation (P$\Delta$E) is an equation or a system of equations involving
unknown functions that depend on more than one independent discrete variables (see e.g. \cite{HJN2016}).
For example, suppose $\uu_{n,m}=(u^1_{n,m}, \dots, u^d_{n,m})$ is a vector-valued
function, depending on two integer independent variables $n, m\in \ZZ$, then an autonomous
P$\Delta$E for $\uu$ would take the following form:
\begin{equation}
    \mathrm{Q}\left(\uu_{n, m}, \dots, \uu_{n+N, m+M}\right)=0,
    \quad \forall n, m\in\ZZ, \label{pde-1}
\end{equation}
where $\mathrm{Q}=(Q^1, \dots, Q^d)$ is a vector-valued function.

The goal of this paper is to introduce a new method to analyze P$\Delta$Es of the above form from the perspective of
partial differential equations (PDEs). To this end, we introduce the following concept.
\begin{dfn}\label{dfn-tf}
    An evolutionary PDE
    \begin{equation}
        \frac{\p u^{\alpha}}{\p t}(x,t)=q^{\alpha}\left(\uu(x,t), \frac{\p \uu}{\p x}(x,t),
        \dots;\e\right), \quad \alpha=1, \dots, d, \label{pde-2}
    \end{equation}
    is called a tautological flow of the P$\Delta$E \eqref{pde-1}, if for any solution
    $\uu(x,t)$ of \eqref{pde-2}, the following assignment
    \begin{equation}
        \uu_{n+k,m+l}=\uu\left(x+k\,\e, t+l\,\e\right),\quad \forall k=0, \dots, N;\ l=0, \dots, M
        \label{pde-3}
    \end{equation}
    gives a solution of \eqref{pde-1}.
\end{dfn}

Many discrete integrable systems possess tautological flows.
\begin{emp}
    Consider the following discrete KdV equation \cite{H1977-I}:
    \begin{equation}
        v_{n,m+1}-v_{n+1,m}=\frac{1}{v_{n,m}}-\frac{1}{v_{n+1,m+1}}. \label{dkdv}
    \end{equation}
    We will show later that it possesses a tautological flow of the form:
    \begin{align}
        v_t= &\, \frac{v^2+1}{v^2-1}v_x+\e^2\left(
        \frac{v^2 \left(v^2+1\right)}{3 \left(v^2-1\right)^3}v_{xxx}\right.\notag            \\
             &\, \left.-\frac{2 v^3 \left(v^2+3\right)}{\left(v^2-1\right)^4}v_x\,v_{xx}
        +\frac{2 v^2 \left(v^4+6 v^2+1\right)}{\left(v^2-1\right)^5}v_x^3
        \right)+\cdots \label{dkdv-tf}
    \end{align}
    Here we denote $v_t=\frac{\p v}{\p t}$, $v_x=\frac{\p v}{\p x}$, $v_{xx}=\frac{\p^2 v}{\p x^2}$,
    and so on.
\end{emp}

From the above example, we see that the tautological flow is usually not truncated,
so we need to treat the right-hand side of the equation \eqref{pde-2} as a formal power
series of $\e$, that is
\begin{align}
    \frac{\p u^{\alpha}}{\p t}= &\, A^{\alpha}_{\beta}\left(\uu\right)u^{\beta}_x+
    \e \left(B^{\alpha}_{\beta}\left(\uu\right)u^{\beta}_{xx}
    +C^{\alpha}_{\beta\gamma}\left(\uu\right)u^{\beta}_x u^{\gamma}_x\right)\notag              \\
    &\, +\e^2 \left(D^{\alpha}_{\beta}\left(\uu\right)u^{\beta}_{xxx}
    +E^{\alpha}_{\beta\gamma}\left(\uu\right)u^{\beta}_x u^{\gamma}_{xx}
    +F^{\alpha}_{\beta\gamma\delta}\left(\uu\right)u^{\beta}_x u^{\gamma}_x u^{\delta}_x
    \right)+\cdots, \label{pde-4}
\end{align}
where $\alpha=1, \dots, d$, and the Einstein summation convention is adopted.

A systematic study of integrable systems of the form \eqref{pde-4} was initiated by Boris A. Dubrovin
and the third author of the present paper in the seminal work \cite{DZ2001}. In recent years, they
and their collaborators have made significant contributions to this field.
When $d=1$, which is called the scalar case, an integrability criterion and a Hamiltonian criterion
were developed in \cite{LZ2006} and \cite{LWuZ2008}. For systems of the form \eqref{pde-4} which possess
semisimple bihamiltonian structures, a systematic classification program and its relation to semisimple
cohomological field theories were established in a series of papers
\cite{LZ2005, DLZ2006, LZ2011, LZ2013, DLZ2018, LWaZ2021, VBCIH-1, VBCIH-2, VBCIH-3, LWaZ-LV}.
By using these results, we can prove properties of tautological flows \eqref{pde-2},
and obtain interesting conclusions about the original P$\Delta$E \eqref{pde-1}.

Why do we consider the equation like \eqref{dkdv-tf} which has a more complicated form comparing with the original equation \eqref{dkdv}?
Our basic idea is similar to certain approaches used in the fields of partial differential equations and number theory.
For example, when finding smooth solutions of an elliptic partial differential equation, researchers first consider
the problem in Sobolev spaces, and use the completeness of these spaces to prove the existence of weak solutions,
then they apply various embedding theorems to obtain solutions with the expected regularity (see e.g. \cite{GT2001}).
Similarly, in number theory, when studying certain arithmetic problems over the ring $\ZZ$ of integers, people first
transfer the problem into the one over the ring $\ZZ_p$ of $p$-adic integers, then obtain certain weak solutions by using
the completeness of $\ZZ_p$, and manage to go back to the original problem (see e.g. \cite{SerreGTM7}). The approach
we take here is similar. For example, the equation \eqref{dkdv} belongs to a rational function field, which lacks
completeness, while the tautological flow \eqref{dkdv-tf} belongs to the ring of differential polynomials, which
is complete (see the next section). The completeness of the differential polynomial ring permits us to prove some
classification theorems, which enable us to prove properties about the original equation \eqref{dkdv}. In fact,
the topology on the differential polynomial ring is essentially the same as the topology on the ring $\ZZ_p$.

In this paper we will focus mainly on some concrete examples to illustrate our basic idea,
since it is currently difficult to obtain general results for the most general form \eqref{pde-1}.
Our principle in selecting these examples is simplicity rather than novelty. Therefore, some of the
conclusions may already be known or can be easily derived by using traditional methods. What we want to
demonstrate is the typical usage of the tautological flow method, rather than specific conclusions
about these examples.

Apart from the discrete KdV equation \eqref{dkdv}, another interesting example that will be considered in this paper
is the following equation,
\begin{equation}
    \hat{u}^{++}\left(\hat{u}-u^{+}\right)^+\left(\hat{u}^{+}\hat{u}-u^+u\right)=
    u\left(\hat{u}-u^{+}\right)\left(\hat{u}^{+}\hat{u}-u^+u\right)^+, \label{dqkdv}
\end{equation}
where $\bullet^+=\bullet_{n+1,m}$, $\hat{\bullet}=\bullet_{n,m+1}$, and so on. As far as we know, this equation currently
does not have a well-known name. We derive it from a discrete symmetry of the $q$-KdV equation (see below), so we name it
\emph{the discrete $q$-KdV equation}. After the submission of this paper, an anonymous reviewer pointed out to us that this
equation has appeared in \cite{BPPS2001} in a slightly different form, and he also showed us an important connection of this
equation with \eqref{dkdv} by comparing the Lax pair of \eqref{dkdv} (see \cite{HJN2016} or \cite{Sh2000}) and the Lax operator
of \eqref{dqkdv} (see \eqref{qkdv-lax}): if $v=v_{n,m}$ is a solution to \eqref{dkdv}, then
\begin{equation}
    u=\pm\left(v+\frac{1}{v^+}\right) \label{dkdv-miura}
\end{equation}
gives a solution to \eqref{dqkdv}. So the relationship between these two equations is similar to the one between
the mKdV equation and the KdV equation, and the above relation is actually a Miura type transformation (see the next section).

Now let us explain why we are interested in the equation \eqref{dqkdv}, and give some interesting applications of the tautological flow method.

In 1985, Boris A. Kupershmidt studied a discrete analogue of the KP hierarchy \cite{K1985}. The Lax operator
he considered takes the form
\begin{equation}
    L_K=\LL^{\beta}+q_0(n)\LL^{\beta-\alpha}+q_1(n)\LL^{\beta-2\alpha}+\cdots, \label{lax-LK}
\end{equation}
where $\LL$ is the shift operator $\LL(f)(n)=f(n+1)$, $q_i\ (i=0, 1, \dots)$ are unknown functions, and $\alpha$, $\beta$
are two positive integers. It is shown in \cite{K1985} that when $\alpha$, $\beta$, and $q_i$'s satisfy certain conditions,
the hierarchy possesses one, two, or three Hamiltonian structures.
In 1991, Frank W. Nijhoff and his collaborators introduced the lattice Gel'fand-Dikii hierarchy \cite{NPCQ1992} (see also \cite{CNP1991,PNGR1991}),
whose Lax operator reads
\begin{equation}
    L_{NPCQ}=-\Sigma_{\eta}^N+\Sigma_{\eta}^{N-1} X^{(1)}+\cdots+\Sigma_{\eta} X^{(N-1)}+\Delta,
\end{equation}
where $\Sigma_\eta$, $X^{(i)}$, and $\Delta$ are certain matrices playing similar roles as $\LL$ and $q_i$.
Their main results include the discrete symmetry of the above operator
and  a Hamiltonian structure. In 1996, Edward V. Frenkel and Nicolai Yu. Reshetikhin introduced
the $q$-deformed $N$-KdV hierarchy \cite{FR1996}. The Lax operator they considered takes the form
\begin{equation}
    L_{FR}=\LL_q^N+u_1(z)\LL_q^{N-1}+\cdots+u_{N-1}(z)\LL_q+\mu, \label{lax-LFR}
\end{equation}
where $\LL_q$ is the $q$-difference operator $\left(\LL_q(f)\right)(z):=f(q z)$, $u_i\ (i=1, \dots, N-1)$ are unknown functions,
and $\mu\in\CC$ is a constant. Note that if we introduce new variables $x=\log z$, $\e=\log q$, and pull back $f(z)$ to the chart of $x$,
then we have $f(q z)=g(x+\e)$, where $g=f\circ \exp$, and
\[\left(\LL_q(f)\right)(z)=f(q z)=g(x+\e)=\left(\LL(g)\right)(x),\]
so there is no essential difference between the $q$-difference operator $\LL_q$ and the usual shift operator $\LL$
(here the assignment \eqref{pde-3} is used to relate the independent variables $x$ and $n$).

The operator $L_{FR}$ is a particular case of $L_K$. According to \cite{K1985}, the integrable hierarchy
defined by the Lax operator $L_{FR}$ should possess only one Hamiltonian structure. This fact was also confirmed in \cite{NPCQ1992}.
Nonetheless, Frenkel and Reshetikhin managed to find out a second Hamiltonian structure.
The key point is that, the discrete independent variable $n$ in \eqref{lax-LK} is replaced by
the continuous independent variable $z$ in \eqref{lax-LFR}, and the whole hierarchy is embedded
into a larger space. What happens here is just like what we have seen in the theory of
partial differential equations or number theory: equations that have no solution in a smaller space may become solvable
in a larger space, although the new solution may only exist in a weak sense.
This example can be seen as the starting point of our idea of the tautological flow technique.

Our initial goal was to find bihamiltonian structures of P$\Delta$Es and use the bihamiltonian cohomology techniques
which we developed previously to study their properties. Since the main purpose of this paper is to introduce
a new method, rather than to discuss specific equations, the examples we choose should be as simple as possible.
Therefore, we will restrict ourselves to the $N=2$ case of the $q$-deformed $N$-KdV hierarchy,
which is also known as the $q$-KdV hierarchy. The Lax operator of the $q$-KdV hierarchy is given by
\begin{equation}
    L(u)=\LL^2+u \LL+\mu,\label{qkdv-lax}
\end{equation}
here we replace the $q$-difference operator $\LL_q$ by the usual one $\LL$, since they are essentially the same.
The $k$-th flow of this hierarchy, which describes the $k$-th isospectral deformation of $L$, is given by
the Lax equation
\begin{equation}
    \frac{\p L}{\p t_k}=\left[\left(L^{k-\frac{1}{2}}\right)_+,\ L\right],
    \quad k=1, 2, \dots \label{qkdv-flows}
\end{equation}
It is shown in \cite{K1985} that $\left[\frac{\p}{\p t_k}, \frac{\p}{\p t_l}\right]=0$
for arbitrary $k, l\in\NN$, so these flows form an integrable hierarchy.

According to \cite{NPCQ1992}, a discrete symmetry $u\mapsto\hat{u}$ of the operator \eqref{qkdv-lax} can be defined as
\begin{equation}
    \hat{L}=M\,L\,M^{-1}, \label{dqkdv-LM}
\end{equation}
where $\hat{L}=L\left(\hat{u}\right)=\LL^2+\hat{u}(z) \LL+\mu$, $M=\LL+b$, and $b$ is a function to be determined.
The equation \eqref{dqkdv-LM} is equivalent to the following ones:
\begin{equation}
    \hat{u}\,b^+=u\,b,\quad \hat{u}+b^{++}=u^++b. \label{dqkdv-ub}
\end{equation}
The first equation of \eqref{dqkdv-ub} yields
\[b^+=\frac{u}{\hat{u}}b, \quad b^{++}=\frac{u^+}{\hat{u}^+}\frac{u}{\hat{u}}b,\]
then the second equation of \eqref{dqkdv-ub} implies that
\[b=\frac{\hat{u}-u^+}{\hat{u}^+\hat{u}-u^+u}\hat{u}^+\hat{u}.\]
By eliminating $b$ in the first equation of \eqref{dqkdv-ub}, we obtain the discrete $q$-KdV equation \eqref{dqkdv}.

Now we have continuous flows \eqref{qkdv-flows} and a discrete symmetry $u\mapsto \hat{u}$ defined by \eqref{dqkdv},
then it is interesting to ask whether they mutually commute (see \cite{LW2006}). This is not an easy problem.
One need to verify that the operator $\hat{L}$ satisfies the following equations:
\begin{equation}
    \frac{\p \hat{L}}{\p t_k}=\left[\left(\hat{L}^{k-\frac{1}{2}}\right)_+,\ \hat{L}\right], \quad k=1, 2, \dots \label{dsym-1}
\end{equation}
These equations are equivalent to the ones for the operator $M$:
\begin{equation}
    \frac{\p M}{\p t_k}=\left(\hat{L}^{k-\frac{1}{2}}\right)_+ M-M \left(L^{k-\frac{1}{2}}\right)_+, \quad k=1, 2, \dots \label{dsym-2}
\end{equation}
When $k=1$, one can verify it directly, but it becomes highly nontrivial for larger $k$.
On the other hand, we have the following result.
\begin{thm}[Corollary 3.6. of \cite{LZ2006}] \label{thm-a0}
    All continuous symmetries of the evolutionary PDE
    \begin{align}
        u_t =& f(u)u_x+\e\,\left(f_1(u)u_{xx}+f_2(u)u_x^2\right) \notag \\
        &+\e^2\,\left(f_3(u)u_{xxx}+f_4(u)u_x u_{xx}+f_5(u)u_x^2\right)+\cdots, \mbox{ with } f'(u)\ne 0\label{pde-scalar}
    \end{align}
    mutually commute.
\end{thm}

By using this theorem, we can prove the following result, which is the first application of the tautological flow method.
\begin{thm}\label{thm-a}
    The discrete transformation $u\mapsto \hat{u}$ defined by \eqref{dqkdv} is a symmetry of the whole $q$-KdV hierarchy \eqref{qkdv-flows}.
\end{thm}

It is proved in \cite{LZ2006} that for an evolutionary PDE of the form \eqref{pde-scalar} there exists a quasi-Miura type transformation
(see Section \ref{sec-4} for more details) which transforms all the continuous symmetries of the equation \eqref{pde-scalar} to their leading terms.
It is further shown in \cite{LWuZ2008} that this transformation also transforms Hamiltonian structures of \eqref{pde-scalar} to their leading
terms. These results lead to the following theorem.
\begin{thm}[Theorem 1.2 of \cite{LZ2006}] \label{thm-b0}
    If the evolutionary PDE \eqref{pde-scalar} has a Hamiltonian structure $P$,
    then $P$ is also a Hamiltonian structure for all continuous symmetries of \eqref{pde-scalar}.
\end{thm}

The bihamiltonian structure $(P_1, P_2)$ of the $q$-KdV hierarchy reads (see \cite{FR1996})
\begin{align}
    P_1 & =\LL-\LL^{-1}, \label{ho-p1}                                \\
    P_2 & =u\,\frac{1-\LL}{1+\LL}\,u+\mu(\LL-\LL^{-1}), \label{ho-p2}
\end{align}
and the flow \eqref{qkdv-flows} can be written as
\[\frac{\p u}{\p t_k}=P_1\left(\frac{\delta H_k}{\delta u}\right)
    =P_2\left(\frac{\delta H_{k-1}}{\delta u}\right),\]
where $H_k$'s are the conserved quantities
\[H_k=\frac{2}{2k+1}\int \res \left(L^{k+\frac{1}{2}}\right)\dd x, \quad k=0, 1, 2, \dots\]
By using the above theorem, we can prove the following result, which is the second application of the tautological flow method  (see \cite{S2003}).
\begin{thm} \label{thm-b}
    The discrete symmetry $u\mapsto \hat{u}$ defined by \eqref{dqkdv} preserves the bihamiltonian structure $(P_1, P_2)$.
\end{thm}

In the above two applications, we only use the existence of tautological flows of the equation \eqref{dqkdv},
rather than their explicit expressions.
This existence is proved through the equations \eqref{dqkdv-ub}, which relies on the fact that we have known
the relationship between the Lax operator \eqref{qkdv-lax} and the equation \eqref{dqkdv}.
However, when faced with a new P$\Delta$E, we may not know its Lax pair. In this case, we can still approximately calculate
its tautological flow to a certain order in $\e$, and use this approximated tautological flow to find out the continuous symmetries,
Hamiltonian structures, and bihamiltonian structures of the original equation.
Thanks to the helpful comments from the anonymous reviewer, we learned that the equations \eqref{dkdv} and \eqref{dqkdv}
are related via the Miura type transformation \eqref{dkdv-miura}. This important connection enables us to
demonstrate an application of approximated tautological flows and bihamiltonian cohomology techniques to the study of P$\Delta$Es.

Suppose we are only given the two equations \eqref{dkdv} and \eqref{dqkdv} themselves, without any \emph{a priori}
knowledge of their Lax pairs, Hamiltonian structures, and so on, then what can we do with them?
We summarize our tautological flow method in the following steps:
\begin{enumerate}
    \item Calculate the tautological flows of the equations \eqref{dkdv} and \eqref{dqkdv} to a relatively high order in $\e$.
    \item Calculate the quasi-Miura type transforms of these tautological flows by using the algorithm given in \cite{LZ2006}.
    \item Search for bihamiltonian structures of these tautological flows by using the algorithm given in \cite{LWuZ2008}.
          It turns out that both tautological flows admit semisimple bihamiltonian structures.
    \item Compute the leading terms and the central invariants of these two bihamiltonian structures.
    \item By comparing the results, we see that these two bihamiltonian structures have the same leading terms and
          central invariants, so they must be equivalent via a Miura type transformation (Corollary 1.11 of \cite{DLZ2006}).
    \item Find a Miura type transformation that transforms the tautological flow of \eqref{dkdv} to that of \eqref{dqkdv}.
\end{enumerate}
The Miura type transformations between the two bihamiltonian hierarchies, if they exist, are not unique, 
and we can choose the following one as the final result
\[u=\left(v+\frac{1}{v}\right)+\e\left(-\frac{v_x}{v^2}\right)+\e^2\left(\frac{2v_x^2-v\,v_{xx}}{2\,v^3}\right)+\cdots\]
which coincides with the Miura type transformation \eqref{dkdv-miura} with the positive sign.

The above procedure seems more complicated than the Lax pair method. However, all the algorithms,
criteria, and classification theorems involved in each of the above-mentioned six steps were developed for general
purposes, and do not require specific information about the equations in consideration. In the most important
fifth step, we only need to compare a finite number of invariants to determine whether these two bihamiltonian
structures are equivalent. Therefore, in this sense, the above procedure is ``simpler'' than the Lax pair method.

In the above fifth step, we used the triviality of the second-order bihamiltonian cohomology of a semisimple bihamiltonian structure.
It would be very interesting to find applications of the higher-order bihamiltonian cohomologies and the variational bihamiltonian cohomologies
\cite{VBCIH-1,VBCIH-2,VBCIH-3} in the study of P$\Delta$Es.


The rest of the paper is organized as follows.
In Section \ref{sec-2}, we provide a brief introduction to the formal calculus of differential polynomials
and prove Theorem \ref{thm-a} and Theorem \ref{thm-b}. Then in Section \ref{sec-3}, we show how to find
approximated tautological flows of P$\Delta$Es, and discuss some issues that may arise in the procedure of
finding such flows; we also propose an interesting conjecture (Conjecture \ref{cnj-a})
concerning the existence of higher-order terms in tautological flows. In Section \ref{sec-4}, we take the discrete
$q$-KdV equation as an example to illustrate how approximated tautological flows and their quasi-triviality
transformations can be used to search for continuous symmetries and bihamiltonian structures of a P$\Delta$E.
We also apply this method to the discrete KdV equation, and explain how to find the hidden relation between this equation and the discrete $q$-KdV equation by using the bihamiltonian cohomology method. Finally, in the
last section, we present two collections of interesting problems that will guide our further research.
We also provide an explicit expression for the tautological flows of the discrete $q$-KdV equation up to
$\e^6$-terms in Appendix \ref{app-tf}.

\section{The ring of differential polynomials}\label{sec-2}

\subsection{Definitions and properties}

In this section, we give a brief introduction to the formal calculus of differential polynomials,
which is mainly developed in \cite{LZ2011} (see also \cite{L2018}).

Let $U$ be a contractible open set in $\RR^d$ with coordinate $\uu=\left(u^1, \dots, u^d\right)$.
We denote by $\A_0$ the ring of smooth functions on $U$, and introduce infinitely-many \emph{jet variables}
$\{u^{\alpha, s}\mid \alpha=1, \dots, d;\ s=1, 2, \dots\}$,
then \emph{the ring of differential polynomials} on $U$ is defined as
\begin{equation}
    \A=\A(U)=\A_0[[u^{\alpha, s}\mid \alpha=1, \dots, d;\, s=1, 2, \dots]].
\end{equation}
Here the formal power series ring $\A_0[[\dots]]$ is completed with respect to the following gradation,
which is called \emph{the standard gradation},
\[\deg f=0, \mbox{ if } f\in \A_0;\quad \deg u^{\alpha,s}=s.\]
The homogeneous component of $\A$ with respect to the standard gradation is denoted by $\A_p\ (p\ge 0)$.
In particular, the ring $\A_0$ of smooth functions on $U$ is just the degree zero component of $\A$.
Furthermore, we denote $u^{\alpha,0}=u^\alpha$, then a differential polynomial in $\A$ is
just a formal function of variables $\{u^{\alpha, s}\mid \alpha=1, \dots, d;\ s=0, 1, 2, \dots\}$.

We would like to include a formal parameter $\e$ in $\A$ to indicate degrees of monomial terms.
More precisely, for a differential polynomial $f\in\A$, we usually write it as
\[f=f_0+\e f_1+\e^2 f_2+\cdots, \quad \mbox{where } f_p\in\A_p,\]
and if $g\in \A_{\ge 1}$, we usually write it as
\[g=g_1+\e g_{2}+\e^2 g_{3}+\cdots, \quad \mbox{where } g_p\in\A_p,\]
and so on.

There is an important derivation $\p$ of $\A$
\[\p=\sum_{s\ge 0} u^{\alpha, s+1}\frac{\p}{\p u^{\alpha,s}},\]
which is called \emph{the translation derivation} of $\A$.
This derivation maps $\A_p$ into $\A_{p+1}$, so its exponential map is well-defined. We denote
\begin{equation}
    \LL=e^{\e \p}=\sum_{k \ge 0}\frac{1}{k!} \e^k \p^k \label{dpr-shift}
\end{equation}
and call it \emph{the shift operator} of $\A$.

It is shown in Lemma 9 of \cite{L2018} that evolutionary PDEs of the following form:
\[u^{\alpha}_t=X^\alpha,\quad X^\alpha \in \A,\quad \alpha=1, \dots, d\]
are in one-to-one correspondence with continuous derivations of $\A$ that commute with the translation derivation $\p$.
A derivation satisfying these conditions always takes the following form:
\[D_X=\sum_{s\ge 0}\p^s\left(X^\alpha\right)\frac{\p}{\p u^{\alpha, s}},
\quad \mbox{where } X=\left(X^1, \dots, X^d\right)\in \left(\A\right)^d.\]
Here the term $\p^s\left(X^\alpha\right)$ means the result of the action of the operator $\p^s$
on the differential polynomial $X^\alpha \in \A $.
We will call $D_X$ (or $X$ for short) an \emph{evolutionary vector field} on $U$.
All evolutionary vector fields form a Lie algebra, which is denoted by $\E$.

The standard gradation on $\A$ can be extended to $\E$.
When $X\in\E_{\ge1}$, we also have $D_X\left(\A_p\right)\subseteq \A_{p+1}$, so the exponential map of $D_X$ is also well-defined.
We denote
\[\LL_X=e^{\e D_X}=\sum_{k \ge 0}\frac{1}{k!} \e^k D_X^k,\]
and call it \emph{the shift operator along the evolutionary vector field} $X\in\E_{\ge1}$.

Now we can give a new definition of tautological flows based on the notion of differential polynomials.
\begin{dfn}
    An evolutionary vector field $X\in \E_{\ge 1}$ is called a tautological flow of the P$\Delta$E \eqref{pde-1},
    if the following assignment
    \[\uu_{n+k,m+l}=\LL^k\LL_X^l(\uu),\quad \forall k=0, \dots, N;\ l=0, \dots, M\]
    gives a solution of \eqref{pde-1}.
\end{dfn}

The map $\uu\mapsto \hat{\uu}=\uu_{n,m+1}=\LL_X\left(\uu\right)$ gives an automorphism of
the topological differential $\RR$-algebra $\A$. We introduce the following more general notion.

\begin{dfn}
    A Miura type transformation is a transformation of the following form
    \begin{equation}
        \uu\mapsto \hat{\uu}=\mathrm{G}_0+\e \, \mathrm{G}_1+\e^2 \, \mathrm{G}_2+\cdots, \label{miura}
    \end{equation}
    where $\mathrm{G}_0$ is a diffeomorphism from $U$ to another open set $\hat{U}\subseteq \RR^d$,
    and $\mathrm{G}_k\in\left(\A_k\left(U\right)\right)^d$.
    A Miura type transformation is always an isomorphism from $\A(U)$ to $\A\left(\hat{U}\right)$.
    In particular, if $\hat{\uu}=\mathrm{G}_0$, it is called a Miura type transformation of the first type,
    while if $\mathrm{G}_0=\mathrm{Id}_{U}$, it is called a Miura type transformation of the second type.
\end{dfn}

The map $\uu\mapsto \hat{\uu}=\LL_X\left(\uu\right)$ which is defined by an evolutionary vector field $X\in\E_{\ge 1}$
is just a Miura type transformation of the second type. Conversely, we have the following lemma.

\begin{lem}[Lemma 22 in \cite{L2018}]\label{lemma22}
    For any Miura type transformation of the second type $\uu\mapsto \hat{\uu}$, there exists a unique evolutionary vector field
    $X\in\E_{\ge 1}$ such that $\hat{\uu}=\LL_X\left(\uu\right)$.
\end{lem}

Therefore, we give the following definition.

\begin{dfn}
    If $X\in\E_{\ge 1}$ is a tautological flow of the equation \eqref{pde-1}, then the corresponding Miura type transformation
    of the second type $\uu\mapsto \hat{\uu}=\LL_X\left(\uu\right)$ is called a formal solution to the equation \eqref{pde-1}.
\end{dfn}

According to Lemma \ref{lemma22}, tautological flows are in one-to-one correspondence with formal solutions.

\begin{emp}
    The tautological flow \eqref{dkdv-tf} of the discrete KdV equation \eqref{dkdv} corresponds to the following formal solution:
    \begin{align}
        \hat{v} = &\, v+\e\left(\frac{v^2+1}{v^2-1}\right)v_x
        +\e^2\left(\frac{\left(v^2+1\right)^2}{2 \left(v^2-1\right)^2}v_{xx}\right.\notag                         \\
                  &\, \left.-\frac{4 \left(v^3+v\right)}{\left(v^2-1\right)^3}v_x^2\right)
        +\e^3\left(\frac{\left(v^2+1\right) \left(v^4+4 v^2+1\right)}{6 \left(v^2-1\right)^3}v_{xxx}\right.\notag \\
                  &\, \left.-\frac{2 v \left(4 v^4+9 v^2+3\right)}{\left(v^2-1\right)^4}v_{xx}v_x
        +\frac{8 v^6+42 v^4+28 v^2+2}{\left(v^2-1\right)^5}v_x^3
        \right)+\cdots\label{dkdv-fs}
    \end{align}
\end{emp}

\subsection{The $q$-KdV hierarchy and its discrete symmetry}\label{sec-dqkdv}

Now let us consider the $q$-KdV hierarchy and its discrete symmetry.
We take $d=1$, and define the ring $\A$ of differential polynomials and the Lie algebra $\E$ of evolutionary vector fields as above.
Then one can show that all the flows \eqref{qkdv-flows} are elements of $\E$.
For example, the first flow of the $q$-KdV hierarchy, which is known as the $q$-KdV equation, takes the following form:
\begin{equation}
    \frac{\p u}{\p t_1}=u\left(1-\LL\right)a, \label{qkdv-1}
\end{equation}
where $a$ is a function satisfying the relation $\left(1+\LL\right)a=u$.
From the point of view of differential polynomial ring $\A$, the operator $1+\LL$ is invertible
\[\left(1+\LL\right)^{-1}=\frac{1}{1+e^{\e\p}}
    =\frac{1}{2}+\sum_{n\ge 1}(1-4^n)\frac{B_{2n}}{(2n)!}
    \left(\e\,\p\right)^{2n-1},\]
where $B_{2n}$ are the Bernoulli numbers, so we have $a\in \A$, and the $q$-KdV equation becomes
\begin{equation}
    \frac{\p u}{\p t_1}=u\,\frac{1-\LL}{1+\LL}\,u
    =\sum_{n\ge 1}2(1-4^n)\frac{B_{2n}}{(2n)!}
    \e^{2n-1}\left(u\,u^{(2n-1)}\right) \in \A.\label{qkdv-first}
\end{equation}

On the other hand, the condition \eqref{dqkdv-ub} implies that
\[\left(\LL-1\right)((u-b^+)b)=0.\]
Note that the kernel of the operator $\LL-1$ on $\A$ is just $\RR$, so there exists a constant $\lambda\in\RR$ such that
\begin{equation}
    u=b^++\frac{\lambda}{b}. \label{dqkdv-ubb}
\end{equation}
Let $b=b_0+\e\,b_1+\e^2\,b_2+\cdots$, then it is easy to obtain two possible $b_0$:
\begin{equation}
    b_{0,\pm}=\frac{1}{2} \left(u\pm\sqrt{u^2-4 \lambda}\right), \label{dqkdv-b0}
\end{equation}
and all $b_i\ (i\ge 1)$ are differential polynomials of $b_0$, so $b\in\A$.
After fixing a choice of $b$, we can represent the shifted $u$ as
\begin{equation}
    \hat{u}=b+\frac{\lambda}{b^+}.\label{dqkdv-hubb}
\end{equation}
The constant $\lambda$ is called the spectral parameter of the
discrete symmetry $u\mapsto \hat{u}$.

From the equations \eqref{dqkdv-ubb}-\eqref{dqkdv-hubb} we see that the transformation $u\mapsto\hat{u}$ is indeed
a Miura type transformation of the second type, so it admits a tautological flow. Note that this transformation depends
on the choice of the spectral parameter $\lambda$ and the sign $\pm$ in \eqref{dqkdv-b0}.

\vskip 1em

\begin{prfn}{Theorem \ref{thm-a}}
    We prove the $k=1$ case of \eqref{dsym-2} first, which is equivalent to the following equation:
    \begin{equation}
        b_{t_1}=b(\hat{a}-a), \mbox{ where } \hat{a}=a^++b-b^+. \label{qkdv-b1}
    \end{equation}
    Note that the relation \eqref{dqkdv-ubb} is a Miura type transformation, which is invertible in the ring of differential polynomials,
    so we only need to show that the equation \eqref{qkdv-b1} implies \eqref{qkdv-1}:
    \begin{align*}
        u_{t_1} =& b^+_{t_1}-\frac{\lambda}{b^2}b_{t_1}\\
        =& b^+(a^{++}-a^++b^+-b^{++})+(b^+-u)(a^+-a+b-b^+)\\
        =& b^+(a^{++}-b^{++}-a+b)-u(a^+-a+b-b^+) \\
        =& b^+(u^{+}-b^{++}+b-u)+u(a-a^+-b+b^+) \\
        =& b^+\hat{u}+u(a-a^+-b)=u(a-a^+).
    \end{align*}
    Here we have used the relations \eqref{dqkdv-ub}, \eqref{dqkdv-ubb}, and $u=a+a^+$.

    Let $X\in\E$ be the tautological flow of the discrete transformation $u\mapsto\hat{u}$.
    The $k=1$ case of \eqref{dsym-1} or \eqref{dsym-2} implies that $[\frac{\p}{\p t_1},X]=0$. According to Theorem \ref{thm-a0},
    the tautological flow $X$ also commutes with all the other flows of \eqref{qkdv-flows}, so $u\mapsto\hat{u}$ gives a discrete symmetry
    of the whole hierarchy. 
\end{prfn}

\vskip 1em

\begin{prfn}{Theorem \ref{thm-b}}
We have seen that the tautological flow $X$ is a continuous symmetry of $\frac{\p}{\p t_1}$, then Theorem \ref{thm-b0} implies
that the bihamiltonian structure $(P_1, P_2)$ is also a bihamiltonian structure of $X$, so the discrete transformation
$u\mapsto\hat{u}$ preserves $(P_1, P_2)$. 
\end{prfn}

\section{How to find approximated tautological flows}\label{sec-3}

In this section, we will explain how to find approximated tautological flows of a P$\Delta$E of the form \eqref{pde-1}.
In principle, we only need to set
\[X=X_1+\e X_2+\e^2 X_3+\cdots ,\quad \mbox{where } X_p\in\left(\A_p\right)^d,\]
substitute $\uu_{n+k,m+l}=\LL^k\LL_X^l(\uu)$ for $k=0, \dots, N$ and $l=0, \dots, M$
into \eqref{pde-1}, then find $X_1, X_2, \dots$ one by one.
However, many issues may arise in this process, and we will explain how to deal with them.

\subsection{The zeroth order}

If we take $\e=0$ in \eqref{pde-3}, then $\uu_{n+k, m+l}=\uu(x,t)$ for arbitrary $k$ and $l$, and the equation \eqref{pde-1} becomes
\begin{equation}
    \mathrm{Q}_0:=\mathrm{Q}\left(\uu, \dots, \uu\right)=0. \label{zeroth-cond}
\end{equation}
This is a necessary condition for the existence of a tautological flow.

In practice, many P$\Delta$Es do not satisfy the condition $\mathrm{Q}_0=0$.
However, we can preprocess these equations to make them satisfy it.

We first simplify equation \eqref{pde-1}. By introducing more unknown functions, we may assume $M=1$
without loss of generality. If we denote $\uu_n=\uu_{n,m}$, $\hat{\uu}_n=\uu_{n,m+1}$,
then the equation \eqref{pde-1} becomes
\begin{equation}
    \mathrm{Q}\left(\uu_n, \dots, \uu_{n+N};\ \hat{\uu}_n, \dots, \hat{\mathrm{u}}_{n+N}\right)=0. \label{simple-pde-1}
\end{equation}

\begin{dfn}
    A Miura type transformation
    \[\uu\mapsto \hat{\uu}=G=\mathrm{G}_0+\e \mathrm{G}_1+\e^2 \mathrm{G}_2+\cdots\]
    is called a \emph{generalized formal solution} to the P$\Delta$E \eqref{simple-pde-1}, if the assignment
    \begin{equation}
        \uu_{n+k}=\LL^k(u), \quad \hat{\uu}_{n+k}=\LL^k\left(G\right),\quad k=0, \dots, N
    \end{equation}
    gives a solution to \eqref{simple-pde-1}.
\end{dfn}

Similarly, by taking $\e=0$, we obtain the following necessary condition for the existence of a generalized formal solution:
\begin{equation}
    \mathrm{Q}\left(\uu, \dots, \uu;\ \mathrm{G}_0\left(\uu\right), \dots, \mathrm{G}_0\left(\uu\right)\right)=0. \label{gene-zeroth-cond}
\end{equation}
This condition is an implicit function equation for the unknown diffeomorphism $\mathrm{G}_0$. If this diffeomorphism exists,
then we can take the transformation of unknown functions $\tilde{\uu}=\mathrm{G}_0^{-1}\left(\hat{\uu}\right)$, such that the Miura type
transformation $\uu\mapsto\tilde{\uu}$ is of the second type, then it corresponds to a tautological flow.

\subsection{The leading order}\label{sec-leading}

In this subsection, we consider the leading order approximation of a P$\Delta$E of the form \eqref{pde-1}
satisfying the zeroth order condition $\mathrm{Q}_0=0$.

We only consider the scalar case, that is $d=1$. The $d>1$ cases are similar but more complicated.
The upper indices $\alpha=1, \dots, d$ of $u$ or $Q$ are omitted, since we always have $\alpha=1$.
Suppose the tautological flow takes the following form
\begin{equation}
    u_t=f(u)u_x+O(\e), \label{tf-order-1}
\end{equation}
then we have
\[u_{n+k,m+l}=u(x+k\,\e,t+l\,\e)=u+\e\left(k+l\,f(u)\right)u_x+O(\e^2),\]
and the equation \eqref{pde-1} becomes
\[\e\left(\sum_{k=0}^N\sum_{l=0}^M\left(k+l\,f(u)\right)\left.\frac{\p Q}{\p u_{n+k, m+l}}\right|_{\e=0}\right)+O(\e^2)=0,\]
here we used the condition $Q_0=0$. We denote
\[K_1=\sum_{k=0}^N\sum_{l=0}^M k \left.\frac{\p Q}{\p u_{n+k, m+l}}\right|_{\e=0},\quad
    L_1=\sum_{k=0}^N\sum_{l=0}^M l \left.\frac{\p Q}{\p u_{n+k, m+l}}\right|_{\e=0},\]
then the function $f(u)$ satisfies a simple equation
\begin{equation}
    K_1+f(u) L_1=0. \label{fK1L1}
\end{equation}
If $L_1\ne 0$, we have
\[f(u)=-K_1/L_1.\]

\begin{emp}\label{dkdv-leading}
    For the discrete KdV equation \eqref{dkdv}, the function $Q$ reads
    \[Q=v_{n,m+1}-v_{n+1,m}-\frac{1}{v_{n,m}}+\frac{1}{v_{n+1,m+1}},\]
    so we have
    \[K_1=-1-\frac{1}{v^2}, \quad L_1=1-\frac{1}{v^2},\quad \mbox{and }f=\frac{v^2+1}{v^2-1}.\]
\end{emp}

However, there are many P$\Delta$Es which do not satisfy the condition $L_1\ne 0$.
For example, the discrete $q$-KdV equation \eqref{dqkdv} satisfies $K_1=0$, $L_1=0$.
Furthermore, there exists P$\Delta$Es satisfying $K_1\ne 0$ and $L_1=0$, so they do not have tautological flows.

\begin{emp}
    Consider the following P$\Delta$E:
    \[u\,u^+ + u\,\hat{u} + u\,\hat{u}^+ + u^+\,\hat{u}^+ + 2\,\hat{u}\,\hat{u}^+=6\,u^+\,\hat{u}.\]
It satisfies the condition $Q_0=0$, $L_1=0$, and $K_1=u$, so it does not admit a tautological flow.
\end{emp}

We assume from now on that the function $Q$ satisfies the conditions
\[Q_0=0, \quad K_1=0, \quad L_1=0,\]
and consider the second order Taylor expansion of the equation \eqref{pde-1}.
The tautological flow takes the following form:
\begin{equation}
    u_t=f(u)u_x+\e\left(f_2(u)u_{xx}+f_{11}(u)u_x^2\right)+O(\e^2), \label{tf-order-2}
\end{equation}
so we have
\begin{align*}
    u_{xt} = & f(u)u_{xx}+f'(u)u_x^2+O(\e),        \\
    u_{tt} = & f(u)^2u_{xx}+2f(u)f'(u)u_x^2+O(\e).
\end{align*}
Then by using the Taylor formula 
\[u_{n+k,m+l}=u+\e \left(k\,u_x+l u_t\right)+\frac{\e^2}2\left(k^2 u_{xx}+2\, k\, l u_{xt}+l^2 u_{tt}\right)+O(\e)^3,\]
one can obtain the following Taylor expansion of the equation \eqref{pde-1}:
\[Q=Q_0+\e\, Q_1\, u_x+\frac{\e^2}{2}\left(Q_2 u_{xx}+Q_{11} u_x^2\right)+O(\e^3),\]
where $Q_0=0$, $Q_1=K_1+L_1 f=0$, and
\begin{align}
     & Q_2= \sum_{k=0}^N\sum_{l=0}^M\left(k+l\,f\right)^2\left.\frac{\p Q}{\p u_{n+k, m+l}}\right|_{\e=0},                          \\
     & Q_{11}= \sum_{k=0}^N\sum_{l=0}^M\left(2\,k\,l f'+2\,l^2\,f\,f'\right)\left.\frac{\p Q}{\p u_{n+k, m+l}}\right|_{\e=0}\notag \\
     & +\sum_{k_1=0}^N\sum_{l_1=0}^M\sum_{k_2=0}^N\sum_{l_2=0}^M\left(k_1+l_1\,f\right)\left(k_2+l_2\,f\right)
    \left.\frac{\p^2 Q}{\p u_{n+k_1, m+l_1}\p u_{n+k_2, m+l_2}}\right|_{\e=0}.
\end{align}
Note that the new unknown functions $f_2$ and $f_{11}$ do not appear in these equations,
since their coefficients contain $L_1$, which is assumed to be zero.
So the equations $Q_2=0$ and $Q_{11}=0$ give an overdetermined system for the function $f(u)$. If it has a solution, then we find the leading term
of the tautological flow \eqref{tf-order-2}. Otherwise, there is no tautological flows for the original P$\Delta$E.

\begin{emp}
    Consider the following P$\Delta$E:
   \begin{align*}
         & u^2+(u^+)^2+\hat{u}^2+(\hat{u}^+)^2+u\,\hat{u}^+                                 \\
         & \quad =u\,u^+ + u\,\hat{u} + u^+\, \hat{u} + u^+\, \hat{u}^+ + \hat{u}\, \hat{u}^+,
    \end{align*}
    It satisfies the condition $Q_0=0$, $Q_1=0$, and
    \[Q_2=u\,f(u),\quad Q_{11}=u\,f'(u)+f(u)^2-f(u)+1.\]
    The equation $Q_2=0$ implies $f(u)=0$, then $Q_{11}=1\ne 0$, so the above  P$\Delta$E does not admit a tautological flow.
\end{emp}

If a P$\Delta$E satisfies the condition that $Q_0$, $K_1$, $K_1$, $Q_2$, $Q_{11}$ all vanishes, then we need to consider the third order approximation.
This is what happens in the case of the discrete $q$-KdV equation \eqref{dqkdv}. Suppose
\[u_t=f(u)u_x+\e(\cdots)+\e^2(\cdots)+O(\e^3),\]
here the terms in $(\cdots)$ do not appear in the final equation, so we omit their forms. Then the discrete $q$-KdV equation \eqref{dqkdv} becomes
\[\e^3\left(2\,u\left(u\,f'(u)+f(u)^3-f(u)\right)u_x^3\right)+O(\e^4)=0.\]
Thus, the unknown function $f$ satisfies the following ODE:
\[u\,f'(u)+f(u)^3-f(u)=0,\]
whose general solution reads
\begin{equation}
    f(u)=\pm \frac{u}{\sqrt{u^2+C}}.\label{dqkdv-f}
\end{equation}
Here the arbitrary constant $C$ is related to the spectral parameter $\lambda$ that appears in \eqref{dqkdv-ubb} and \eqref{dqkdv-hubb}
via the relation $C=-4\lambda$. So the leading term $f(u)$ has the same degree of freedom with the leading term of the auxiliary functions $b$
(see \eqref{dqkdv-b0}).

\subsection{The higher orders}

Suppose we have found the leading term $u_t=f(u)u_x+O(\e)$ of a tautological flow of the equation \eqref{pde-1},
we set the higher order terms as
\begin{align*}
    u_t = &\, f(u)u_x+\e\left(f_2(u)u_{xx}+f_{11}(u)u_x^2\right)                                \\
          & \, +\e^2\left(f_3(u)u_{xxx}+f_{21}(u)u_{xx}u_x^2+f_{111}(u)u_x^3\right)+\cdots,
\end{align*}
where $f_2,\ f_{11},\ \dots$ are unknown functions to be determined, then consider the higher order Taylor expansion of the equation \eqref{pde-1}.

\begin{emp}
    We have found the leading term of the tautological flow of the discrete KdV equation \eqref{dkdv} in Example \ref{dkdv-leading},
then one can obtain
    \[f_2=0,\quad f_{11}=0,\quad f_3=\frac{v^2 \left(v^2+1\right)}{3 \left(v^2-1\right)^3},\quad
        f_{21}=-\frac{2 v^3 \left(v^2+3\right)}{\left(v^2-1\right)^4},\dots\]
    just like what we have shown in \eqref{dkdv-tf}.
\end{emp}

For the discrete $q$-KdV equation \eqref{dqkdv}, the leading term has been given in \eqref{dqkdv-f}, then
one can obtain that
\begin{align*}
     & f_2=0,\quad f_{11}=0,                                                                                          \\
     & f_3=-\frac{1}{12}u\,f'(u),\quad f_{21}=-\frac{1}{6}u\,f''(u),\quad f_{111}=-\frac{1}{24}u\,f'''(u),\dots
\end{align*}
Higher order terms can be found in Appendix \ref{app-tf}.

In this calculation, it seems that there is no any obstruction for the existence of higher order terms, and the solution is always unique,
so we have the following conjecture.

\begin{cnj}\label{cnj-a}
    Suppose $d=1$, if one can find a function $f(u)$ such that $f'(u)\ne 0$
    and $u_t=f(u)u_x+O(\e)$ gives a first order approximated tautological flow of \eqref{pde-1}, then there exists a unique
    tautological flow of \eqref{pde-1} with the following form:
    \[u_t=f(u)u_x+X,\quad \mbox{where } X\in\A_{\ge2}.\]
\end{cnj}

The nondegenerate condition $f'(u)\ne 0$ is very useful. It is introduced in \cite{LZ2006}, and we will show in the next section
how to use the results of \cite{LZ2006} and \cite{LWuZ2008} to study P$\Delta$Es satisfying this condition.

\section{Quasi-triviality and its applications} \label{sec-4}

The tautological flow \eqref{dqkdv-tf-full} of the discrete $q$-KdV equation given in Appendix \ref{app-tf} seems to be complicated.
However, each of its monomial has the same pattern:
\[\left(C_{p,q,d_0,d_1,d_2,\dots, d_m} f^{(p)}(u)u^{d_0}u_x^{d_1}u_{xx}^{d_2}\cdots u_{mx}^{d_m}\right) \e^q ,\]
where $C_{p,q,d_0,d_1,d_2,\dots, d_m}$ are constants, and
\[\sum_{i=0}^m d_i=p+1, \quad \sum_{i=0}^m i\,d_i=q+1.\]
This pattern suggests that this PDE might be a formal symmetry of another simpler PDE (see \cite{LZ2006}).

According to the results of \cite{LZ2006}, every PDE of the form
\[u_t=f(u)u_x+X,\quad f'(u)\ne0, \quad X\in\A_{\ge2}\]
admit a reducing transformation, which is a quasi-Miura type transformation that transforms the above equation to its leading term.
A quasi-Miura type transformation is a transformation similar to \eqref{miura}, in which the functions $G_1$, $G_2$, $\dots$ can be
rational or logarithmic functions of the jet variables $\{u^{\alpha, s}\mid \alpha=1, \dots, d;\ s=1, 2, \dots\}$.
In particular, by using the algorithm given in \cite{LZ2006}, one can obtain the reducing transformation
\begin{align}
    u= &\, \bar{u}+\epsilon ^2 \left(\frac{\bar{u} \bar{u}_{xx}^2}{24 \bar{u}_x^2}-\frac{\bar{u} \bar{u}_{3x}}{24 \bar{u}_x}\right)+
        \epsilon ^4 \left(\frac{\bar{u}^2 \bar{u}_{6x}}{1152 \bar{u}_x^2}+\frac{3
        \bar{u} \bar{u}_{5x}}{640 \bar{u}_x}-\frac{\bar{u} \bar{u}_{3x}^2}{72 \bar{u}_x^2}+\frac{\bar{u}^2 \bar{u}_{xx}^5}{18 \bar{u}_x^6}\right. \notag                                                  \\
       &\, -\frac{37 \bar{u} \bar{u}_{xx}^4}{1152 \bar{u}_x^4}-\frac{41 \bar{u}^2
        \bar{u}_{5x} \bar{u}_{xx}}{5760 \bar{u}_x^3}-\frac{73 \bar{u}^2 \bar{u}_{3x} \bar{u}_{4x}}{5760 \bar{u}_x^3}+
        \frac{17 \bar{u}^2 \bar{u}_{4x} \bar{u}_{xx}^2}{480 \bar{u}_x^4}-\frac{13 \bar{u}
        \bar{u}_{4x} \bar{u}_{xx}}{640 \bar{u}_x^2}\notag                                                                                                             \\
       &\, \left.-\frac{35 \bar{u}^2 \bar{u}_{3x} \bar{u}_{xx}^3}{288 \bar{u}_x^5}+
       \frac{71 \bar{u} \bar{u}_{3x} \bar{u}_{xx}^2}{1152 \bar{u}_x^3}+\frac{19 \bar{u}^2
        \bar{u}_{3x}^2 \bar{u}_{xx}}{384 \bar{u}_x^4}\right)+\cdots \label{dqkdv-qm}
\end{align}
of the tautological flow \eqref{dqkdv-tf-full}. It transforms the tautological flow \eqref{dqkdv-tf-full} to the following equation
\[\bar{u}_t=f(\bar{u})\bar{u}_x.\]
Then, according to the principles introduced in \cite{LZ2006}, we apply the inverse transformation of \eqref{dqkdv-qm}
to a simpler symmetry
\[\bar{u}_{s_1}=\bar{u}\,\bar{u}_x.\]
The result should be a simpler symmetry of \eqref{dqkdv-tf-full}, which reads
\begin{align}
    u_{s_1}= &\, u\,u_x-\frac{\e^2}{12}u\,u_{3x}+\frac{\e^4}{120}u\,u_{5x}-\frac{17\e^6}{20160}u\,u_{7x}\notag         \\
             & \, +\frac{31\e ^8}{362880}u\,u_{9x}-\frac{691\e^{10} }{79833600}u\,u_{11x}+\cdots \label{dqkdv-ss}
\end{align}
After a careful observation, it turns out that
\[u_{s_1}=2\,u\frac{\LL-1}{\LL+1}u+O(\e^{12}),\]
which is proportional to the first flow of the $q$-KdV equation \eqref{qkdv-first}.
If we apply the inverse transformation of \eqref{dqkdv-qm} to other symmetries
\[\bar{u}_{s_k}=\bar{u}^k\,\bar{u}_x,\quad k=3, 5, 7, \dots\]
we can obtain other flows of the $q$-KdV hierarchy \eqref{qkdv-flows}  (up to a linear combination).

Note that in the above calculation, we do not use any
\emph{a priori} knowledge of the $q$-KdV equation \eqref{qkdv-first}. Everything, including the tautological flow \eqref{dqkdv-tf-full},
its reducing transformation \eqref{dqkdv-qm}, its simpler symmetry \eqref{dqkdv-ss}, comes from the original equation \eqref{dqkdv}.
Therefore, the above calculation demonstrates a powerful analyzing method for studying P$\Delta$Es.

Furthermore, we can use the reducing transformation \eqref{dqkdv-qm} and the results of \cite{LWuZ2008} to search for Hamiltonian structures
of the discrete $q$-KdV equation \eqref{dqkdv}. We only need to apply the inverse transformation of \eqref{dqkdv-qm} to a
simple Hamiltonian structure
\[\{\bar{u}(x),\bar{u}(y)\}_\phi=2\phi(\bar{u}(x))\delta'(x-y)+\phi'(\bar{u}(x))\bar{u}'(x)\delta(x-y),\]
where $\phi(\bar{u})$ is a function to be determined. The result
\[\{u(x),u(y)\}_\phi=2\phi(u(x))\delta'(x-y)+\phi'(u(x))u'(x)\delta(x-y)+\cdots\]
contains a lot of terms that are not differential polynomials.
By requiring the vanishing of these ``bad'' terms, we obtain some ODEs for the function $\phi(u)$, whose
general solution reads
\[\phi(u)=C_1+C_2\,u^2, \quad \forall C_1, C_2\in\RR.\]
(The calculation, which is completed by a computer program, is quite long, so we have to omit the details here.)
Then it is easy to see that the coefficients of $C_1$ and $C_2$ in $\{\bullet,\,\bullet\}_\phi$ gives the bihamiltonian structure
$(P_1, P_2)$ \eqref{ho-p1}-\eqref{ho-p2} of the $q$-KdV hierarchy (up to a linear combination).

The notion of central invariants of a semisimple bihamiltonian structure is introduced in \cite{DLZ2006}.
For a $d$-dimentional semisimple bihamiltonian structure, its central invariants $c_1(\lambda_1), \dots, c_d(\lambda_d)$
are $d$ one-variable functions of the corresponding canonical coordinates $\lambda_1, \dots, \lambda_d$.
There is an explicit formula of central invariants (see the equation (1.49) of \cite{DLZ2006}).
A full explanation to this formula is too long to be presented here, so we have to omit the details, and give the result only.
The leading terms of the bihamiltonian structure $(P_1, P_2)$ \eqref{ho-p1}-\eqref{ho-p2} in the canonical coordinate $\lambda=-\frac{u^2}{4}$ read
\begin{align}
    \{\lambda(x),\lambda(y)\}_1 =& -2\lambda(x)\delta'(x-y)+\cdots, \label{pb1}\\
    \{\lambda(x),\lambda(y)\}_2 =& -2\lambda(x)^2\delta'(x-y)+\cdots, \label{pb2}
\end{align}
and this bihamiltonian structure has the central invariant $c(\lambda)=\frac{1}{24}$. Here we take $\mu=0$ in $P_2$ to eliminate the unnecessary linear combination.

All the above calculation can be applied to the discrete KdV equation \eqref{dkdv}.
For example, the reducing transformation of the tautological flow \eqref{dkdv-tf} reads
\begin{align}
v =& \, \bar{v}+\e^2\left(\frac{\bar{v}(\bar{v}^2+1)}{24 \left(\bar{v}^2-1\right)}
\left(\frac{\bar{v}_{3x}}{\bar{v}_x}-\frac{\bar{v}_{xx}^2}{\bar{v}_x^2}\right)\right.\notag\\
& \, \left.-\frac{5 \bar{v}^2-1}{12 \left(\bar{v}^2-1\right)^2}\bar{v}_{xx}
+\frac{9 \bar{v}^4-2 \bar{v}^2+1}{12 \bar{v} \left(\bar{v}^2-1\right)^3}\bar{v}_x^2\right)+\cdots \label{dkdv-qm}
\end{align}
We can also apply it to a simple symmetry $\bar{v}_{s_1}=\bar{v}\,\bar{v}_x$, and obtain the corresponding flow
$v_{s_1}=v\,v_x+\cdots$. However, in the $q$-KdV case, this flow is hard to recognize, so we omit its explicit expression here.

By using the algorithm given in \cite{LWuZ2008}, one can obtain two Hamiltonian structures
$\{\bullet,\,\bullet\}_{\phi_1}$ and $\{\bullet,\,\bullet\}_{\phi_2}$ of the tautological flow \eqref{dkdv-tf}, where
\[\phi_1(v)=\frac{v^4}{(v^2-1)^2},\quad \phi_2(v)=\frac{v^2(v^4+1)}{(v^2-1)^2}.\]
Note that the bihamiltonian structure can be chosen up to a linear combination, so we choose
\begin{align*}
    \{\bullet,\,\bullet\}_1 =& a_{11}\{\bullet,\,\bullet\}_{\phi_1}+a_{12}\{\bullet,\,\bullet\}_{\phi_2},\\
    \{\bullet,\,\bullet\}_2 =& a_{21}\{\bullet,\,\bullet\}_{\phi_1}+a_{22}\{\bullet,\,\bullet\}_{\phi_2},
\end{align*}
where $a_{ij}\ (i,j=1,2)$ are constants to be determined. Then one can obtain the canonical coordinate
\[\lambda = \frac{a_{21}v^2+a_{22}(1+v^4)}{a_{11}v^2+a_{12}(1+v^4)},\]
and the leading terms of the bihamiltonian structure $\left(\{\bullet,\,\bullet\}_1, \{\bullet,\,\bullet\}_2\right)$ read
\begin{align}
    \{\lambda(x),\lambda(y)\}_1 =& 8\frac{v^4(x) \left(v^2(x)+1\right)^2 (a_{11}a_{22}-a_{12}a_{21})^2
    }{\left(a_{11}v^2(x)+a_{12}(1+v^4(x))\right)^3}\delta'(x-y)+\cdots, \label{pb3}\\
    \{\lambda(x),\lambda(y)\}_2 =& 8\frac{v^4(x) \left(v^2(x)+1\right)^2 (a_{11}a_{22}-a_{12}a_{21})^2
    }{\left(a_{11}v^2(x)+a_{12}(1+v^4(x))\right)^3}\lambda(x)\delta'(x-y)+\cdots. \label{pb4}
\end{align}
Then the central invariant of the bihamiltonian structure can be calculated by using the equation (1.49) of \cite{DLZ2006}
\[c(\lambda) = -\frac{a_{11}v^2+a_{12}(1+v^4)}{96 \,v^2 \,(a_{11}a_{22}-a_{12}a_{21})}.\]
By taking
\[a_{11}=1,\quad a_{12}=0,\quad a_{21}=-\frac12,\quad a_{22}=-\frac14,\]
we obtain
\[\lambda=-\frac{\left(v^2+1\right)^2}{4 v^2}, \quad c(\lambda)=\frac{1}{24},\]
and the leading terms of \eqref{pb3} and \eqref{pb4} become same with \eqref{pb1} and \eqref{pb2}.

Since the bihamiltonian structures of the tautological flows \eqref{dkdv-tf} and \eqref{dqkdv-tf-full} have the same leading
terms and central invariants, they must be equivalent via a certain Miura type transformation, according to the main result of \cite{DLZ2006}.

In fact, the leading term of these Miura type transformations can be obtained from the expression of the canonical coordinate
\[\lambda=-\frac{u^2}{4}=-\frac{\left(v^2+1\right)^2}{4 v^2},\]
which implies that
\[\bar{u}=\pm\left(\bar{v}+\frac{1}{\bar{v}}\right).\]
Then by using the reducing transformation \eqref{dqkdv-qm}, \eqref{dkdv-qm} and their inverse transformation,
one can obtain one of the Miura type transformations that transform \eqref{dkdv-tf} into \eqref{dqkdv-tf-full}. 
Other Miura type transformations can be obtained from the known one by inserting an arbitrary continuous symmetry into its generator.

\section{Concluding Remarks}

In this paper, we propose a new analyzing method for P$\Delta$Es based on PDEs,
namely the tautological flow method. By using this method, we can prove that
the discrete $q$-KdV equation is a discrete symmetry of the $q$-KdV hierarchy and its bihamiltonian structure,
and we also demonstrate how to directly search for continuous symmetries and bihamiltonian
structures of P$\Delta$Es by using the approximated tautological flows and their quasi-triviality transformation.
We believe that the tautological flow method is a powerful analyzing tool and can yield significant
results in the study of other P$\Delta$Es.
In what follows, we list two collections of interesting problems, which will be the focus of our further research.

\paragraph{i)\ On discrete $q$-deformed $N$-KdV equation and its generalization}\mbox{}

Our analysis for the discrete $q$-KdV equation can be generalized to more general $q$-deformed $N$-KdV hierarchies \cite{FR1996}.
For example, when $N=3$, we have the $q$-deformed Boussinesq hierarchy, whose Lax operator reads
\begin{equation}
    L(u,v)=\LL^3+u \,\LL^2+v \,\LL+\mu,\quad \mu\in\CC. \label{qbsq-lax}
\end{equation}
The discrete transformation $L\mapsto \hat{L}=L\left(\hat{u}, \hat{v}\right)$ is also given by
\begin{equation}
    \hat{L}=M\,L\,M^{-1}, \label{qbsq-LM}
\end{equation}
where $M=\LL+b$, and $b$ is a function to be determined. The conditions for $\hat{u}$, $\hat{v}$, and $b$ becomes
\begin{align}
\hat{u}+b^{+++} &= b+u^{+},\label{qbsq-uvb1}\\
\hat{v}+\hat{u} \, b^{++} &= b \, u+v^{+},\label{qbsq-uvb2}\\
\hat{v} \, b^{+} &= b \, v. \label{qbsq-uvb3}
\end{align}
By eliminating $\hat{u}$ and $\hat{v}$ in \eqref{qbsq-uvb1}-\eqref{qbsq-uvb3}, we obtain
\[\left(\LL-1\right)\left(b^{++}\,b^{+}\,b^{}-u\,b^{+}\,b^{}+v\,b^{}\right)=0,\]
so there also exists a spectral parameter $\lambda$ such that
\[b^{++}\,b^{+}\,b^{}-u\,b^{+}\,b^{}+v\,b^{}=\lambda,\]
which implies that $b$ is a differential polynomial of $u$ and $v$.

On the other hand, by eliminating $b$ in \eqref{qbsq-uvb1}-\eqref{qbsq-uvb3}, we obtain two P$\Delta$Es for $\hat{u}$ and $\hat{v}$
\begin{align}
\hat{v}^{+++}\frac{\left(\hat{u}-u^{+}\right)^{+}}{\hat{u}-u^{+}}
&=v\frac{\left(\hat{v}^{++}\hat{v}^{+}\hat{v}^{}-v^{++}v^{+}v^{}\right)^{+}}{\hat{v}^{++}\hat{v}^{+}\hat{v}^{}-v^{++}v^{+}v^{}}, \label{qbsq-uv1}\\
\hat{v}^{++}\left(\hat{u}-u^{+}\right)\left(u\,\hat{v}^{+}\hat{v}^{}-\hat{u}\,v^{+}v^{}\right)
&=\left(\hat{v}-v^{+}\right)\left(\hat{v}^{++}\hat{v}^{+}\hat{v}^{}-v^{++}v^{+}v^{}\right),\label{qbsq-uv2}
\end{align}
which can be named as \emph{the discrete $q$-Boussinesq equations}. Note that a lattice Boussinesq equation has been introduced
in \cite{NPCQ1992}, there may exist a Miura type transformation between these equations.

We believe that the discrete transformation $(u,v)\mapsto \left(\hat{u}, \hat{v}\right)$
defined by the discrete $q$-Boussinesq equations \eqref{qbsq-uv1} and \eqref{qbsq-uv2} gives a discrete symmetry of
the $q$-deformed Boussinesq hierarchy and its bihamiltonian structure (see \cite{FR1996}).
Our proof for the discrete $q$-KdV case (see Theorem \ref{thm-a} and Theorem \ref{thm-b}) depends heavily on the fact
that there is only one unknown function $u$ in \eqref{dqkdv}. For the P$\Delta$Es with more unknown functions like
\eqref{qbsq-uv1} and \eqref{qbsq-uv2}, new techniques need to be developed.

The $q$-deformed $N$-KdV hierarchies is a $q$-deformation of the $N$-KdV hierarchies, which is the Drinfeld-Sokolov hierarchy associated with
the affine Lie algebra of $\hat{A}_{N-1}$ type \cite{DS1985}. There also exist $q$-deformations of the Drinfeld-Sokolov hierarchies associated with
other affine Lie algebras \cite{FRS1998, SS1998}, so it is interesting to consider the discrete symmetries for those general cases
by using the tautological flow method.

\paragraph{ii)\ On classification of discrete integrable systems}\mbox{}

In Subsection \ref{sec-leading}, we point out that the existence of tautological flows is a complicated problem,
even when considering only the leading terms. We present several examples which possess or do not possess tautological flows,
all of them share the following form:
\[Q\left(u,\ u^{+},\ \hat{u},\ \hat{u}^{+}\right)=0,\]
where $Q$ is a polynomial with a certain degree. Note that a polynomial with a given degree and variables
can only contain a finite number of undetermined coefficients, we can establish a classification scheme
to search for all discrete integrable systems that take this form.

More precisely, let $D, N$ be positive integers, and
\begin{equation}
    Q\left(u,\dots, \LL^N\left(u\right),\ \hat{u},\dots, \LL^N\left(\hat{u}\right)\right) \label{classification}
\end{equation}
be a polynomial of degree $D$. According to the discussion given in
Subsection \ref{sec-leading}, we can consider the Taylor expansion of the equation $Q=0$ order by order. Note that
we only have finite coefficients, so such discussion will stop within finite steps. Suppose we have chosen a reasonable leading term,
then according to Conjecture \ref{cnj-a}, the higher order terms of the tautological flow are expected to be uniquely determined.
Then, we can apply the quasi-triviality method developed in \cite{LZ2006} and \cite{LWuZ2008} to analyze
the integrability of the tautological flow with the chosen leading term. The quasi-triviality method can pick up
integrable ones from a family of evolutionary PDE with finitely many undetermined coefficients.
By this point, we have completed the classification of discrete integrable systems with the form \eqref{classification}.

This classification scheme is similar to the ABS (Vsevolod E. Adler, Alexander I. Bobenko, Yuri B. Suris) classification for
discrete integrable systems on quad-graphs \cite{ABS2003}, but the tautological flow method may be able to deal with more general P$\Delta$Es.
We hope that the results of this classification scheme can be comparable to those of the ABS classification.

\vskip 1em

\noindent\textbf{Acknowledgments}
This work was supported by the National Natural Science Foundation of China (NSFC) No.\,12171268 and No.\,11725104.
We thank the anonymous reviewers for their very helpful comments.

\vskip 1em

\appendix

\section{The tautological flows of the discrete $q$-KdV equation} \label{app-tf}

Let $f(u)$ be a function of the form \eqref{dqkdv-f}, then the corresponding tautological flow of the discrete $q$-KdV equation \eqref{dqkdv}
with $u_t=f(u)u_x$ as its leading term is given by
\begin{align}
    u_t = &\, f(u) u_x-\epsilon^2 \left(\frac{1}{24} u f^{(3)}(u) u_x^3+\frac{1}{6} u f''(u) u_{xx} u_x+\frac{1}{12} u f'(u)u_{3x}\right)\notag \\
          & +\epsilon^4 \left(\frac{1}{1152}u^2 f^{(6)}(u) u_x^5
    +\frac{3}{640} u f^{(5)}(u) u_x^5
    +\frac{1}{90} u^2 f^{(5)}(u) u_{xx} u_x^3\right.\notag                                                                                    \\
          & +\frac{23}{1440}u^2 f^{(4)}(u) u_{3x} u_x^2
    +\frac{1}{20} u f^{(4)}(u) u_{xx} u_x^3
    +\frac{31}{1440}u^2 f^{(4)}(u) u_{xx}^2 u_x\notag                                                                                         \\
          & +\frac{1}{80} u^2 f^{(3)}(u) u_{4x} u_x
    +\frac{1}{48} u^2 f^{(3)}(u) u_{3x} u_{xx}
    +\frac{17}{288} u f^{(3)}(u) u_{3x} u_x^2\notag                                                                                           \\
          & +\frac{11}{144} u f^{(3)}(u) u_{xx}^2 u_x
    +\frac{1}{240} u^2 f''(u) u_{5x}
    +\frac{13}{360} u f''(u) u_{4x} u_x\notag                                                                                                 \\
          & \left.+\frac{1}{18} u f''(u) u_{3x} u_{xx}
    +\frac{1}{120} u f'(u) u_{5x}\right)
    -\epsilon^6\left(\frac{5 u f^{(7)}(u)u_x^7}{7168}\right.\notag                                                                            \\
          & +\frac{u^2 f^{(8)}(u) u_x^7}{5120}
    +\frac{u^3 f^{(9)}(u) u_x^7}{82944}
    +\frac{u f^{(6)}(u) u_{xx} u_x^5}{70}\notag                                                                                               \\
          & +\frac{41 u^2 f^{(7)}(u) u_{xx} u_x^5}{8960}
    +\frac{11 u^3 f^{(8)}(u) u_{xx} u_x^5}{34560}
    +\frac{13 u f^{(5)}(u) u_{3x} u_x^4}{512}\notag                                                                                           \\
          & +\frac{4547 u^2 f^{(6)}(u) u_{3x} u_x^4}{483840}
    +\frac{359 u^3 f^{(7)}(u) u_{3x} u_x^4}{483840}
    +\frac{47 u f^{(4)}(u) u_{4x} u_x^3}{1680}\notag                                                                                          \\
          & +\frac{43 u f^{(5)}(u) u_{xx}^2 u_x^3}{640}
    +\frac{1481 u^2 f^{(5)}(u) u_{4x} u_x^3}{120960}+
    \frac{3049 u^2 f^{(6)}(u) u_{xx}^2 u_x^3}{120960}\notag                                                                                   \\
          & +\frac{u^3 f^{(6)}(u) u_{4x} u_x^3}{896}
    +\frac{487 u^3 f^{(7)}(u) u_{xx}^2 u_x^3}{241920}
    +\frac{23 u f^{(4)}(u) u_{3x} u_{xx} u_x^2}{168} \notag                                                                                   \\
          & +\frac{2467 u^2 f^{(5)}(u) u_{3x} u_{xx} u_x^2}{40320}
    +\frac{53 u f^{(3)}(u) u_{5x} u_x^2}{2880}
    +\frac{199 u^2 f^{(4)}(u) u_{5x} u_x^2}{20160}\notag                                                                                      \\
          & +\frac{43 u^3 f^{(5)}(u) u_{5x} u_x^2}{40320}
    +\frac{229 u^3 f^{(6)}(u) u_{3x} u_{xx} u_x^2}{40320}
    +\frac{151 u f^{(3)}(u) u_{3x}^2 u_x}{3456} \notag                                                                                        \\
          & +\frac{151 u f^{(4)}(u) u_{xx}^3 u_x}{2520}
    +\frac{587 u^2 f^{(4)}(u) u_{3x}^2 u_x}{24192}
    +\frac{61 u f^{(3)}(u) u_{4x} u_{xx} u_x}{864}\notag                                                                                      \\
          & +\frac{2357 u^2 f^{(4)}(u) u_{4x} u_{xx} u_x}{60480}
    +\frac{61 u^2 f^{(5)}(u) u_{xx}^3 u_x}{2240}
    +\frac{109 u^3 f^{(5)}(u) u_{3x}^2 u_x}{40320}\notag                                                                                      \\
          & +\frac{29 u^3 f^{(5)}(u) u_{4x} u_{xx} u_x}{6720}
    +\frac{31 u^3 f^{(6)}(u) u_{xx}^3 u_x}{12096}
    +\frac{97 u f''(u) u_{6x} u_x}{15120}\notag                                                                                               \\
          & +\frac{ u^2 f^{(3)}(u) u_{6x} u_x}{224}
    +\frac{u^3 f^{(4)}(u) u_{6x} u_x}{1680}
    +\frac{49 u f^{(3)}(u) u_{3x} u_{xx}^2}{864}\notag                                                                                        \\
          & +\frac{139 u^2 f^{(4)}(u) u_{3x} u_{xx}^2}{4320}
    +\frac{11 u f''(u) u_{4x} u_{3x}}{540}
    +\frac{43 u^2 f^{(3)}(u) u_{4x} u_{3x}}{2880}\notag                                                                                       \\
          & +\frac{ u^3 f^{(4)}(u) u_{4x} u_{3x}}{480}
    +\frac{7 u^3 f^{(5)}(u) u_{3x} u_{xx}^2}{1920}
    +\frac{ u f''(u) u_{5x} u_{xx}}{72}\notag                                                                                                 \\
          & +\frac{29 u^2 f^{(3)}(u) u_{5x} u_{xx}}{2880}
    +\frac{u^3 f^{(4)}(u) u_{5x} u_{xx}}{720}
    +\frac{17 u f'(u) u_{7x}}{20160}\notag                                                                                                    \\
          & \left.+\frac{17 u^2 f''(u) u_{7x}}{20160}
    +\frac{u^3 f^{(3)}(u) u_{7x}}{6720}\right)+\cdots \label{dqkdv-tf-full}
\end{align}

\vskip 1em

\noindent
Department of Mathematical Sciences \\
Tsinghua University, Beijing 100084, P.\,R. China \\

\noindent
Email addresses: \\
caozl19@mails.tsinghua.edu.cn \\
liusq@tsinghua.edu.cn \\
youjin@tsinghua.edu.cn

\end{document}